\documentclass{aa}
\usepackage[varg]{txfonts}
\usepackage[utf8]{inputenc}
\usepackage{natbib}
\usepackage{graphicx}
\usepackage{amsmath}
\usepackage[dvipsnames]{xcolor}
\usepackage{siunitx}
\usepackage{float}
\usepackage[colorlinks]{hyperref}
\usepackage{soul}
\usepackage{siunitx}
\usepackage{caption}
\usepackage{subcaption}

\begin{document}

\title{On the detection of stellar wakes in the Milky Way: a deep learning approach}

\author{
{Sven P\~oder}\inst{1}\inst{2}
\and {Joosep Pata}\inst{1}
\and Mar\'ia Benito\inst{3}
\and Isaac Alonso Asensio\inst{4}\inst{5}
\and Claudio Dalla Vecchia\inst{4}\inst{5}
}

\institute{
National Institute of Chemical Physics and Biophysics (NICPB), R\"avala 10, Tallinn 10143, Estonia \email{sven.poder@kbfi.ee}\label{inst1}
\and
Tallinn University of Technology, Ehitajate tee 5, Tallinn 19086, Estonia\label{inst2}
\and
Tartu Observatory, University of Tartu, Observatooriumi 1, T\~oravere 61602, Estonia \email{mariabenitocst@gmail.com} \label{inst3}
\and
Instituto de Astrofísica de Canarias, c/ Vía Láctea s/n, E-38205, La Laguna, Tenerife, Spain \label{inst4}
\and
Departamento de Astrofísica, Universidad de La Laguna, Av. Astrofísico
Francisco Sánchez s/n, E-38206 La Laguna, Tenerife, Spain\label{inst5}
}

\abstract
    {Due to poor observational constraints on the low-mass end of the subhalo mass function, the detection of dark matter (DM) subhalos on sub-galactic scales would provide valuable information about the nature of DM. Stellar wakes, induced by passing DM subhalos, encode information about the mass (properties) of the inducing perturber and thus serve as an indirect probe for the DM substructure within the Milky Way (MW).}
   {Our aim is to assess the viability and performance of deep learning searches for stellar wakes in the Galactic stellar halo caused by DM subhalos of varying mass.
   }
   {We simulate massive objects (subhalos) moving through a homogeneous medium of DM and star particles, with phase-space parameters tailored to replicate the conditions of the Galaxy at a specific distance from the Galactic center. The simulation data is used to train deep neural networks with the purpose of inferring both the presence and mass of the moving perturber. We then investigate the performance of our deep learning models and identify limitations of our current approach.
   } 
   {We present an approach that allows quantitative assessment of subhalo detectability in varying conditions of the Galactic stellar and DM halos. We find that our binary classifier is able to infer the presence of subhalos in our generated mock datasets, showing non-trivial performance down to a mass of $5 \times  10^7 \rm \, M_\odot$. In a multiple-hypothesis case, we are also able to discern between samples containing subhalos of different mass. By simulating datasets describing subhalo orbits at different Galactocentric distances, we test the robustness of our binary classification model and find that it performs well with data generated from different initial physical conditions. Out of the phase-space observables available to us, we conclude that overdensity and velocity divergence are the most important features for subhalo detection performance.
   
   }
   {}


\maketitle

\section{Introduction}

The standard $\Lambda$CDM scenario succesfully describes the behaviour of dark matter (DM) on extra-galactic scales \citep{einasto2010dark, zavala2019dark}. Studies of structure formation \citep{1990MNRAS.244..214G}, galaxy clustering \citep{10.1093/mnras/stad585}, supernova luminosities \citep{PERIVOLAROPOULOS2022101659} and CMB correlation functions \citep{planck2018} have left little room for deviations from the CDM model on these scales. 

A key prediction of $\Lambda$CDM, that is yet to be confirmed, is the abundance of dark matter subhaloes on sub-galactic scales. In fact, studies of MW-like galaxy simulations show that the subhalo mass function (SHMF), that is the abundance of subhalos per unit mass, follows a power-law well below the galactic scale (e.g. \citealt{AquariusSim}). In the absence of convincing observational evidence for small-scale dark matter clustering below subhalo masses of $\approx 10^9 \rm \, M_\odot$, other alternative DM models (i.e. warm dark matter, self-interacting dark matter, fuzzy dark matter, etc.) are also allowed. These models impose a cutoff in the SHMF below a specific mass threshold and thus change the expected abundance of dark subhaloes orbiting galaxies \citep{Ostdiek_2022, zavala2019dark}. Constraining the low-mass end of the SHMF is therefore an important test of the CDM scenario as deviations from the expected SHMF behaviour could be explained by alternative DM models (e.g. \citealt{2020PhRvD.101j3023B}). This however is not an easy endeavour as subhaloes less massive than $10^{8-9} \rm \, M_\odot$ are not expected to host any stars due to their small mass and reionization effects \citep{10.1093/mnras/stv2597, 10.1093/mnras/staa2698} - they are dark subhaloes.

In recent years, the expected count of low-mass subhalos inside a Milky Way-sized galaxy in the CDM scenario has been revised, although uncertainties remain. \cite{Garrison} report that the inclusion of baryonic physics actually supresses the size of the expected subhalo population when compared to DMO (dark matter only) simulations. The previous work has been improved on by \cite{Barry_2023}, who used the FIRE-2 simulations (described in \citealt{FIRE2_simulations}), and found that at least 20 dark subhalos of mass $> 10^6 \rm \, M_\odot$ should exist within $<= 30 \rm \, kpc$ of the Galactic center. Given that these theoretical predictions are yet to be robustly validated by empirical observation, the Milky Way presents itself as an ideal laboratory for probing the low-mass end of the SHMF. 

Beyond the Local Group, investigating dark subhalos can be effectively pursued by observing the perturbations they impart on strongly lensed images of distant galaxies and quasars (e.g. \citealt{2024arXiv240414487W}). In recent years, deep learning methods have proven to be valuable tools in this endeavor, owing to their effectiveness in image classification tasks (for a thorough overview see e.g. \citealt{varma2020dark} and the references therein). Inside the Galaxy, a promising method to probe the low-mass end of the SHMF involves to search for gaps or density fluctuations in the distribution of cold stellar streams \citep{2024arXiv240519410B}. For example, \cite{Bonaca_2019} studied the interaction of the GD-1 stream with a massive perturber whose mass range they found to be in a range of $10^6 - 10^8 \rm \, M_\odot$. With improved measurement data, the mass detection limit via stream perturbations could be as low as $10^5 \rm \, M_\odot$ \citep{Bovy_2017}. Another method to detect DM substructure in the Galaxy via pulsar timing array measurements, proposed in \cite{Siegel}, promises the detection of even lower masses.  

The current work focuses on the detection of stellar wakes in the MW - arguably the least studied phenomenon for DM detection in the literature. The underlying concept is built upon the notion that when a massive object moves through a field of stars, it experiences dynamical friction \citep{Chandrasekhar} as it perturbs the phase-space of the surrounding stellar medium \citep{1983A&A...117....9M}. Through gravitational interactions with the perturber, stars are pulled toward it and in time cause a relative overdensity opposite to the direction of the perturber's movement (see e.g. \citealt{Weinberg1986}, who described this effect analytically in the context of infalling satellites).
In recent years, there has been growing interest in exploring the effects of dynamical friction-induced wakes as a promising avenue for investigating dark matter substructure. In the work of \cite{Buschmann}, the authors develop an analytical likelihood formalism to use these perturbations in the stellar phase-space to infer the mass of the DM halo passing through the stars. 

A popular testbed for the detection of stellar wakes has been the MW's largest satellite - the Large Magellanic Cloud (LMC) (see e.g. \citealt{2019ApJ...884...51G, 2021ApJ...916...55T, 2022ApJ...933..113R, Foote_2023}). \cite{2022ApJ...933..113R} studied the response of a static Milky Way to the LMC's infall using linear response theory. More recently, \cite{Foote_2023} studied the wake produced by the infall of the LMC using idealised windtunnel simulations in the context of both CDM and FDM. Notably, they observed that the self-gravity of the DM wake amplifies the extent of the stellar wake, particularly for subhalos with masses of the order of $10^{11}\rm M_\odot$. Going beyond simulations, \cite{conroy_all-sky_2021} observed for the first time the density wake trailing behind the orbit of the LMC using data from Gaia's Early Data Release 3. Their work was expanded on by \cite{fushimi2023determination}, who used the wake to estimate the mass of the LMC's DM halo with the method proposed in \cite{Buschmann}. 

In our work, we focus our attention on perturbers less massive than the LMC and thus broaden the scope of \cite{Foote_2023}. Furthermore, we expand on the work in \cite{Buschmann} as we are including the effects of self-gravity in our study of stellar wakes. This study also builds upon our previous research \cite{BAZAROV2022100667}, which demonstrated the discernible impact of dark subhalos on the phase-space distribution of stars in simulated Milky Way-like galaxies. To address the limitations of our previous work, we investigate dark subhalos in the Milky Way using windtunnel simulations, affording greater control over the signal induced by dark subhalos. Much like in \cite{BAZAROV2022100667}, we tackle the problem in a data driven way using machine learning (ML) in lieu of classical likelihood methods. The reason for this is that the latter becomes untractable as simulation complexity increases and even more so in the case of real data with uncertainties.

The structure of the paper is as follows: In Sect. \ref{sec:sim}, we describe the numerical and physical
details of our simulation setup. In Sect. \ref{sec:ml}, we discuss how we generated the mock data and set up our deep learning approach. Section \ref{sec:results} contains the performance results of the models described in the previous section. Section \ref{sec:discussion} outlines the key limitations of the current work and discusses future directions, while Sect.  \ref{sec:conclusions} summarizes the main conclusions.

\section{Windtunnel simulations}
\label{sec:sim}

In the following, we describe our simulations of an extended object orbiting at 30 kpc from the Galactic center. Adopting a spherically symmetric gravitational potential and total mass $M$, this perturber experiences a stationary wind of simulation particles with a bulk velocity $-v$. We note that in the reference frame of the box, the simulation is equivalent to a setting where a perturber with constant velocity $v$ moves through a homogeneous medium of field particles with constant mass density $\rho$ and isotropic Maxwellian velocity distribution. 

\subsection{Perturber setup}

This physical setup is simulated using \texttt{Pkdgrav3}~\citep{2017ComAC...4....2P, 2023MNRAS.519..300A} which is a highly versatile cosmological N-body gravity code. Although generally used to simulate phenomena on cosmological scales, such as large scale structure formation, it can also be used to accurately study the dynamics of systems down to planetesimal scales (see e.g. \citealt{2023MNRAS.519..300A} and the references therein). In our work, we used \texttt{Pkdgrav3} to simulate a massive perturber moving through a homogeneous medium of background particles in a box with equal side lengths of $L=120 \,\rm kpc$ and periodic boundary conditions in all directions (X, Y, Z). The coordinates of the box are defined in the range $x, y, z \in [-60, 60] \rm \, kpc$ and therefore any particle which is at -60 kpc and moving in the -X direction reappears at +60 kpc once it crosses the boundary. The simulation takes place in the rest frame of the perturber which is stationary in the centre of the box at coordinates (0, 0, 0). To simulate the perturber's motion, we introduce a wind of stellar and DM simulation particles moving from right to left with some bulk velocity $-v$ along the X-axis. The magnitude of this velocity was approximated by assuming a circular orbit for the perturber and taking into account the total dynamical mass of the MW enclosed in the region where the Galactocentric distance is $R < 30 \rm \,$ kpc. In the work of \cite{Karukes_2020}, the mass of MW at this range is found to be approximately $3\times10^{11} \rm \, M_\odot$, resulting in a circular orbital speed of $\sim 200 \rm \, km\,s^{-1}$ at 30 kpc. In the FIRE-2 simulations \cite{10.1093/mnras/stad1395}, the tangential velocity of DM subhaloes with masses larger than $10^7 \rm \, M_\odot$ at the radius of 30 kpc from the Galactic center is somewhere closer to 250 $\rm km\,s^{-1}$. With all this in mind, we chose a fiducial perturber velocity of 225 $\rm km\,s^{-1}$ which is in the middle of these estimates.  

Following~\cite{Buschmann}, the perturber is described by a Plummer density profile. This choice allows us to make a clearer comparison between their results and ours. The density of a Plummer sphere as a function of $r$ is given by

\begin{equation}
    \rho(r) = \frac{3{\rm M_{sh}}}{4\pi{R_s}^3}(1+\frac{r^2}{{R_s}^2})^{-5/2},
\end{equation}
where $R_s$ is the Plummer scale radius, $\rm M_{\rm sh}$ is the subhalo total mass and $r$ represents the radial distance from the subhalo's center. In the same way as in ~\cite{Buschmann, vialactea}, we adopt the following equation for the computation of $R_s$,

\begin{equation}
    R_s = 1.62 \rm \, kpc \times \left(\frac{M_{sh}}{10^8 \rm \, M_\odot}\right)^{1/2}.
\end{equation}

We chose to run our simulation with a range of mass options (in addition to simulations with no subhalo present) in order to gauge how our ML model's performance changes with respect to the mass of the perturber. For the purposes of this study, we adopted the following subhalo masses: $5 \times 10^7 \rm \, M_\odot$, $10^8 \rm \, M_\odot$ and $5 \times 10^8 \rm \, M_\odot$. We did not implement subhalos below $5 \times  10^7 \rm \, M_\odot$ as the stellar wakes produced by perturbers smaller than this are not resolved in the simulations. Likewise, subhalos much more massive that $5 \times 10^8 \rm \, M_\odot$ could host dwarf galaxies and are therefore outside the scope of this work.

\subsection{Background particles and initial conditions}

The background star and DM particles were defined in two grids superimposed on each other but shifted in $x$ and $y$ by $L/(2\times512)$. Although this initialisation is not realistic we don't expect any spurious structures to form due to the sufficiently large velocity dispersion of the background particles. 

\begin{figure}[h]
    \centering
        \includegraphics[width=0.4\textwidth]{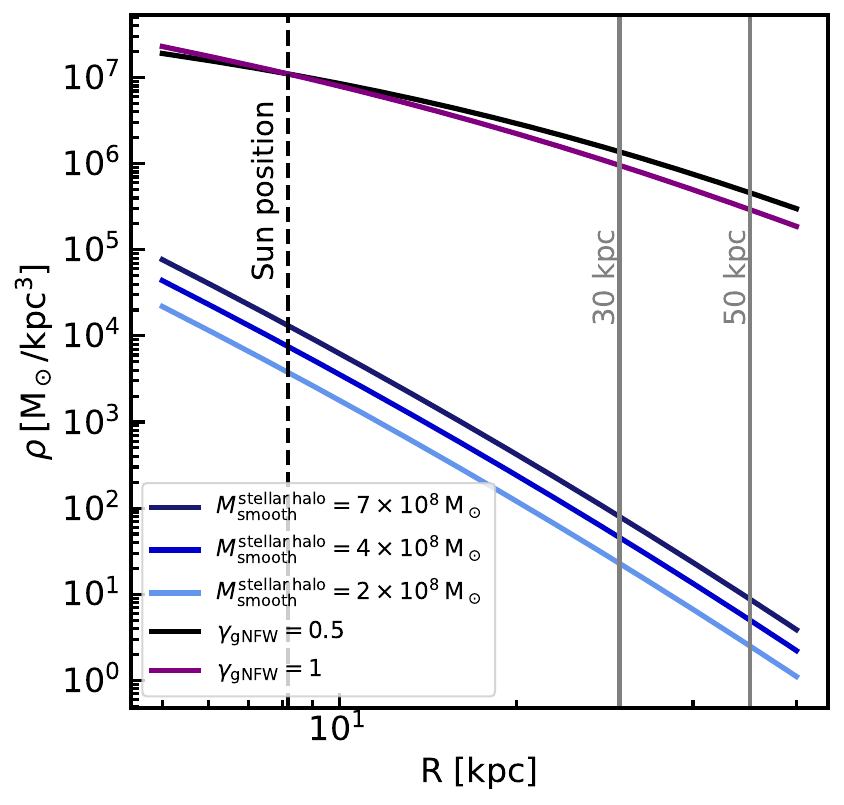}
        \caption{Mass density profile of the virialised stellar and DM haloes in the Milky Way. Both gNFW profiles assume a scale-radius and local DM density values of $R_s=20\,\rm kpc$ and $\rho_0=0.011\,\rm M_\odot/pc^3$~\citep{2021PDU....3200826B}, respectively.}
    \label{fig:mass_profile}
\end{figure}

For the background, we assumed a total mass density of $10^6\,\rm M_\odot\,kpc^{-3}$ for DM and $10^2\,\rm M_\odot\,kpc^{-3}$ for stars. These values roughly match the mass densities of the DM and smooth stellar halo components at 30 kpc from the Galactic center. The stellar halo of the Milky Way, with a mass of approximately 4 to 7 $\times 10^8 \, \rm M_\odot$, comprises of distinct smooth and clumpy components, each contributing roughly equally to the total mass~\citep{2016ARA&A..54..529B, 2019MNRAS.490.3426D}. This study focuses on discerning the influence of dark subhalos within the smooth, virialized portion of the stellar halo, deferring the exploration of detecting stellar wakes within the portion that remains incompletely phase-mixed to future studies. Figure~\ref{fig:mass_profile} shows the mass density profiles of the virialised stellar and DM halos. The former is obtained by fitting the Einasto mass density profile~\citep{1972PhDT........39E} as reconstructed in~\cite{2018ApJ...859...31H} to its total mass. For the total mass we adopt three different values, namely $2\times10^{8}\,\rm M_\odot$, $4\times10^{8}\,\rm M_\odot$ and $7\times10^{8}\,\rm M_\odot$. The DM halo is described by a generalised Navarro-Frenk-White (gNFW) density profile~\citep{1996MNRAS.278..488Z}. From the figure it is clear that the mass density of the DM is always higher than that of the stars, and that this difference increases rapidly with Galactocentric distance. 

In order to simulate the above mentioned ambient densities, we generated $N_{\rm bkg}= 2 \times 512^3$ particles which are divided equally into DM and stellar particle types. The mass values assigned to the two particle types were scaled to satisfy the ratio of total stellar mass to the total DM mass in the Galactic halo. This means that given the number resolution of $2 \times 512^3$, the star particles were assigned a mass $M_{\rm stars}\approx{1.3} \rm \,M_\odot$ whereas the DM particles were initialised with $M_{\rm DM}\approx{1.29 \times 10^4} \rm \,M_\odot$. For the softening of both particle types, we adopted a
widely-used approach in the literature \citep{2017ComAC...4....2P} by setting the softening length to 1/50 of the mean inter-particle
separation such that $\epsilon_{\rm bkg}$=3.72 pc.

We used a 3D isotropic Maxwellian velocity distribution for the velocities of both particles $\Vec{v_{\rm DM}}$ and $\Vec{v_{\rm star}}$. In practice, the velocity components of each Cartesian direction of the DM and star particles were generated by sampling from a 1D Gaussian distribution centred at zero and with standard deviation $\sigma_{\rm DM}$ and $\sigma_{\rm star}$, respectively (see values in Table~\ref{tab:params}). In order to find a reasonable DM particle velocity dispersion ($\sigma_{\rm DM}$), we turned to cosmological simulations of Milky Way-sized galaxies and the reported DM dispersion profiles reported therein. In particular, we looked at studies using data from both the Aquarius Project \citep{10.1111/j.1365-2966.2009.15878.x} and the FIRE-2 simulations \citep{10.1093/mnras/stac966}, and deemed a reasonable DM dispersion at 30 kpc to be $\sigma_{\rm DM}= 200\,\rm km\,s^{-1}$. The choice of the velocity dispersion for the stellar particles ($\sigma_{\rm star}$) was motivated by the Galactocentric velocity dispersion profile of halo stars obtained in \cite{10.1111/j.1365-2966.2012.21639.x}. As we intend to reproduce the physical conditions of the Galactic halo at 30 kpc from the Galactic center, we assumed a value of $\sigma_{\rm star}= 95\,\rm km\,{s^{-1}}$. The physical parameters adopted for this case are summarised in Table~\ref{tab:params}.

\renewcommand{\arraystretch}{1.5}
\begin{table*}
    \caption{Physical parameters adopted in the windtunnel-like N-body simulations for two different locations in the stellar halo.}
    \centering
    \begin{tabular}{l c c c c c c c c}
    \hline\hline
     Case & $r$ [kpc] & $v$ [km/s]& $\rho_{\rm DM}$ $\rm [M_{\odot}/kpc^3]$ & $\sigma_{\rm DM}$ [km/s] & $N_{\rm DM}$ & $\rho_{\rm star}$ $\rm [M_{\odot}/kpc^3]$ & 
     $\sigma_{\rm star}$ [km/s] & $N_{\rm star}$ \\
    \hline
   Case 1 & 30 & 225 & $10^6$ & 200 & $512^3$ & $10^2$ & 95 & $512^3$ \\
    \hline
    Case 2 & 50 & 200 & $10^{5.5}$ & 180& $512^3$ & $10$ & 90 & $512^3$\\
    \hline
    \end{tabular}
    \label{tab:params}
\end{table*}

\subsection{Stellar wakes}
\label{sec:stellarwakes}

\begin{figure*}[t]
    \centering
    \begin{subfigure}[t]{0.4\textwidth}
        \centering
        \includegraphics[width=\linewidth]{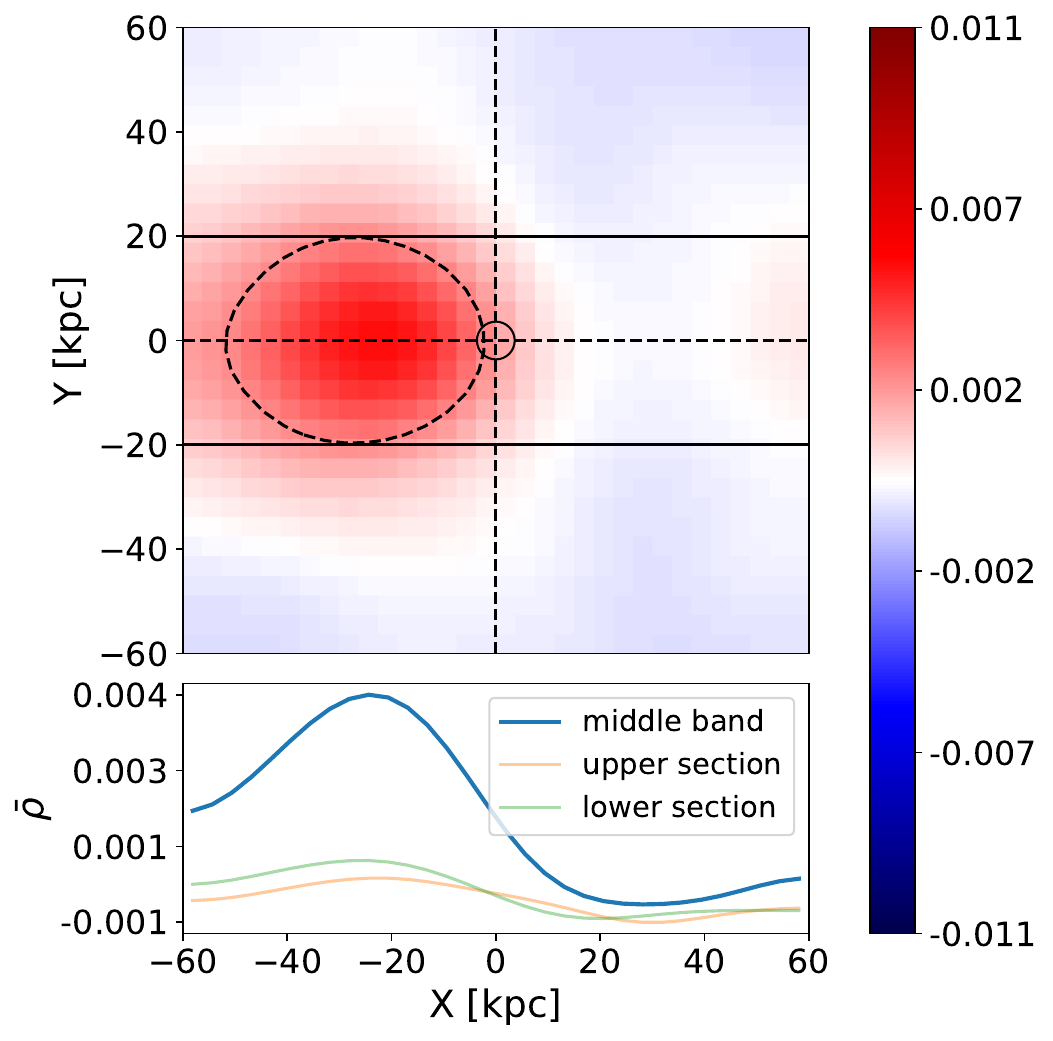} 
        \caption{Overdensity} \label{fig:timing1}
    \end{subfigure}
    \hspace{1cm}
    \begin{subfigure}[t]{0.4\textwidth}
        \centering
        \includegraphics[width=\linewidth]{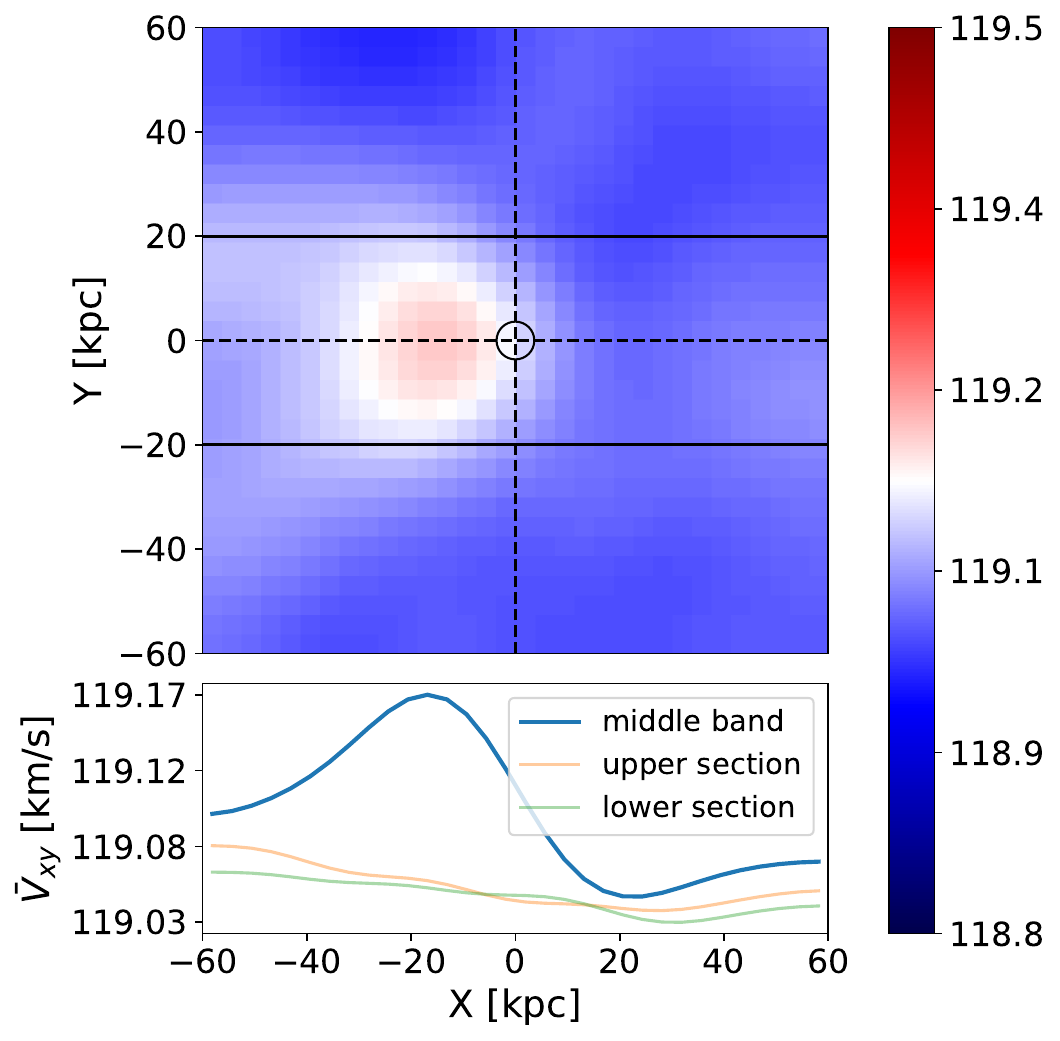} 
        \caption{Mean $X-Y$ speed [km/s]} \label{fig:timing2}
    \end{subfigure}

    \vspace{1cm}

    \begin{subfigure}[t]{0.4\textwidth}
        \centering
        \includegraphics[width=\linewidth]{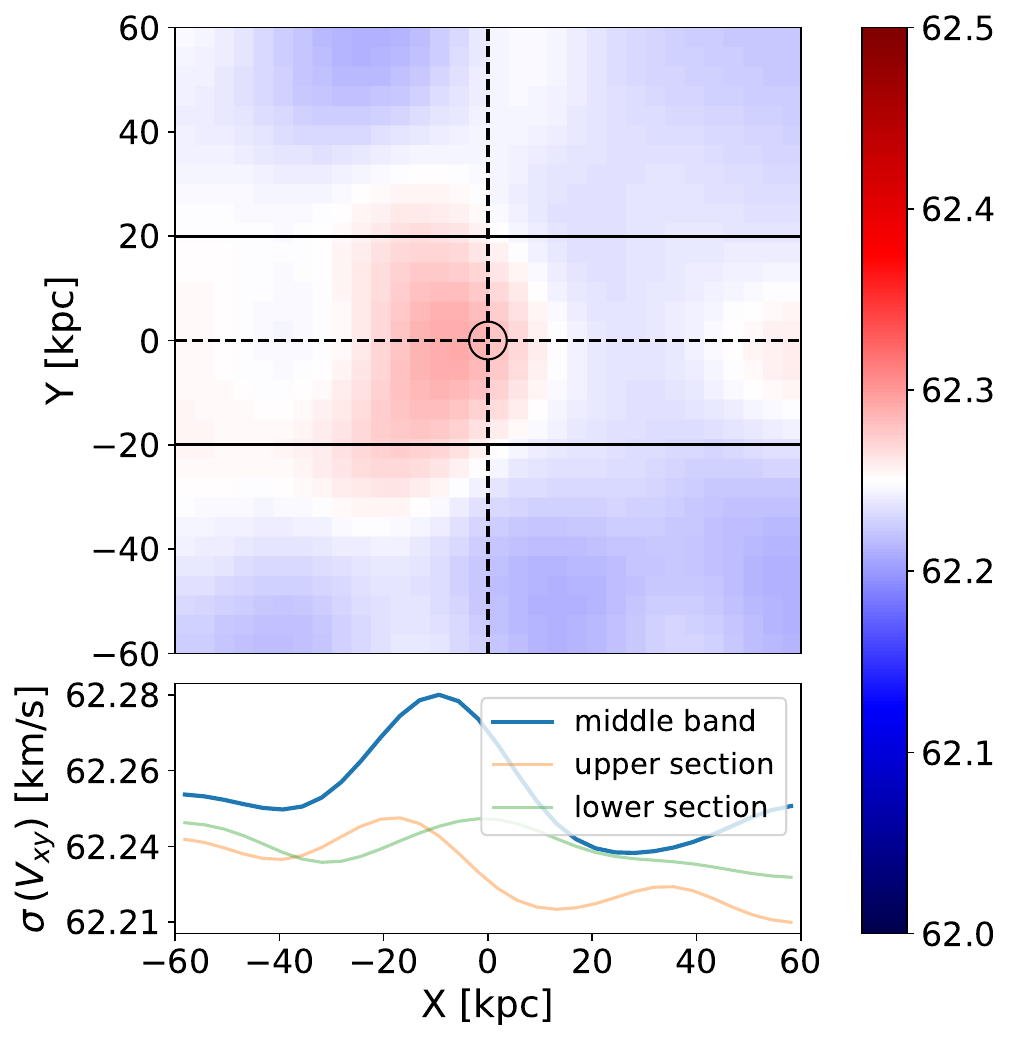} 
        \caption{Dispersion $X-Y$ speed [km/s]} \label{fig:timing3}
    \end{subfigure}
    \hspace{1cm}
    \begin{subfigure}[t]{0.4\textwidth}
        \centering
        \includegraphics[width=\linewidth]{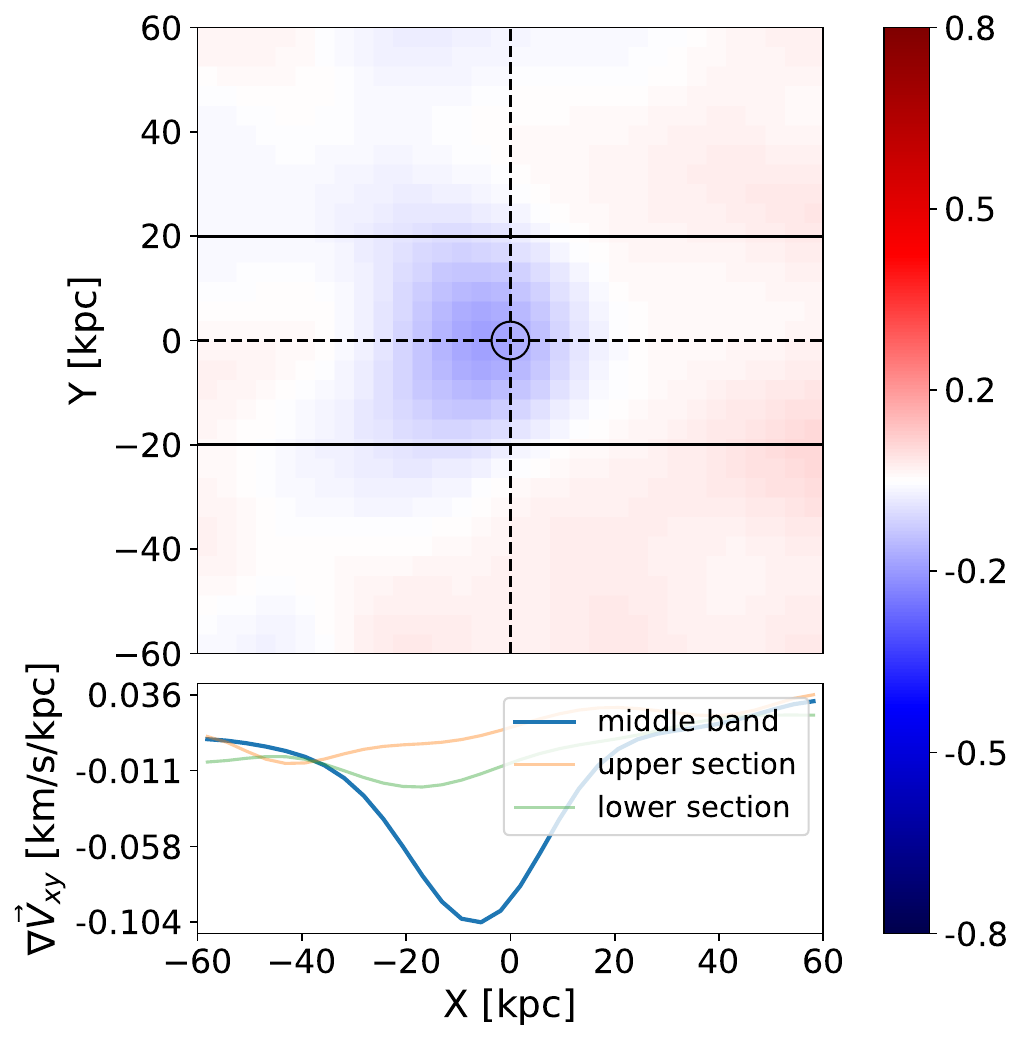} 
        \caption{Divergence $X-Y$ velocity [km/s/kpc]} \label{fig:timing4}
    \end{subfigure}

    \caption{Stellar phase space feature maps, shown in the reference frame of the simulation box (i.e., frame at which the perturber is moving from left to right with an initial speed of 225 km/s), extracted from a simulation with a perturber of mass $5 \times 10^8 \rm \, M_\odot$ after an integration time $t=194.94\,\rm Myr$. Note that in the simulation box's reference frame, the average (unperturbed) 3D velocity of the wind is $\boldsymbol{0}$ and its average (unperturbed) speed in the $X-Y$ plane is 119 km/s. The maps are generated from data contained in a z-slice of $z \in [-20, 20] \, \rm kpc$. Panels (\subref{fig:timing1}) overdensity, (\subref{fig:timing2}) mean speed, (\subref{fig:timing3}) speed dispersion, and (\subref{fig:timing4}) divergence, show the Gaussian-smoothed features projected onto the $X-Y$ plane. Inside the dashed contour of panel (\subref{fig:timing1}), we show the half-max region of the overdensity. Each subfigure includes a lower plot which shows each Y-band's radial profile along the X-axis.
    The perturber is situated in the middle of the histogram with the black circle depicting its scale radius. We see that the wake effects are seen in all four of the phase-space features.}
    
    \label{fig:simout}
\end{figure*}

Figure \ref{fig:simout} shows the star particles of a subhalo simulation with mass  $5 \times 10^8 \rm \, M_\odot$ after an integration time of approximately 195 Myr. At this particular time stamp, the stellar wind has moved a distance of about half the length of the box, giving the wake sufficient time to form. At the same time we took care to avoid snapshots at later times where the simulation particles, having already interacted with the perturber, cross the boundaries on the left and reappear on the right. The reason for this was to prevent any unphysical effects, arising from the wake interacting with itself, manifest in our data. It is because of this, we use simulation snapshots at this particular point in integration time to plot the wake and later generate machine learning (ML) datasets (see Sect. \ref{sec:ml}). The figure is plotted from data which lies in a slice of $z \in [-20, 20] \, \rm kpc$ and it is binned spatially along x and y into 2D histograms with a bin size of 3.75 kpc. For better visibility, we sum the results from ten different simulations with the same subhalo mass and take the mean across these.

In the figure, we show 2D histograms of four phase-space features of the stellar particles: relative overdensity, mean speed, speed dispersion and velocity divergence. For computing the latter three of these features, we use the velocity of the stellar particles in the X-Y plane. Under each figure, we also show how the particular feature varies over the X-coordinate as three profiles which average the quantity in different Y-bands of the simulation box. Instead of raw histograms, we show Gaussian smoothed variants which aim to reduce the overall noise in the figure while preserving the most important features of the wakes. For example, in Fig. \ref{fig:simout}, features in the radial profile of the velocity divergence become much more discernible compared to its unsmoothed counterpart. As also shown in \cite{Foote_2023}, the divergence exhibits a dip behind the formed wake.

The wake is most clearly visible in the upper left panel of Fig. \ref{fig:simout} as an overdense region. The half-max region (i.e. region in which overdensity exceeds half of the maximum, denoted as a dashed line) extends from the middle of the box in the -X direction and is contained between $Y \in [-20, 20]$ kpc. The overdensity inside a particular bin (i,j) is computed with the following equation

\begin{equation}
    \label{eq:overdensity}
    \bar{\rho_{i,j}} = \frac{\rho_{i,j}}{\hat{\rho}} - 1,
\end{equation}
where $\rho_{i,j}$ is the mass density inside bin (i,j) and $\hat{\rho}$ is the average stellar mass density in the simulation box.

\begin{figure}[h]
    \centering
        \includegraphics[width=0.45\textwidth]{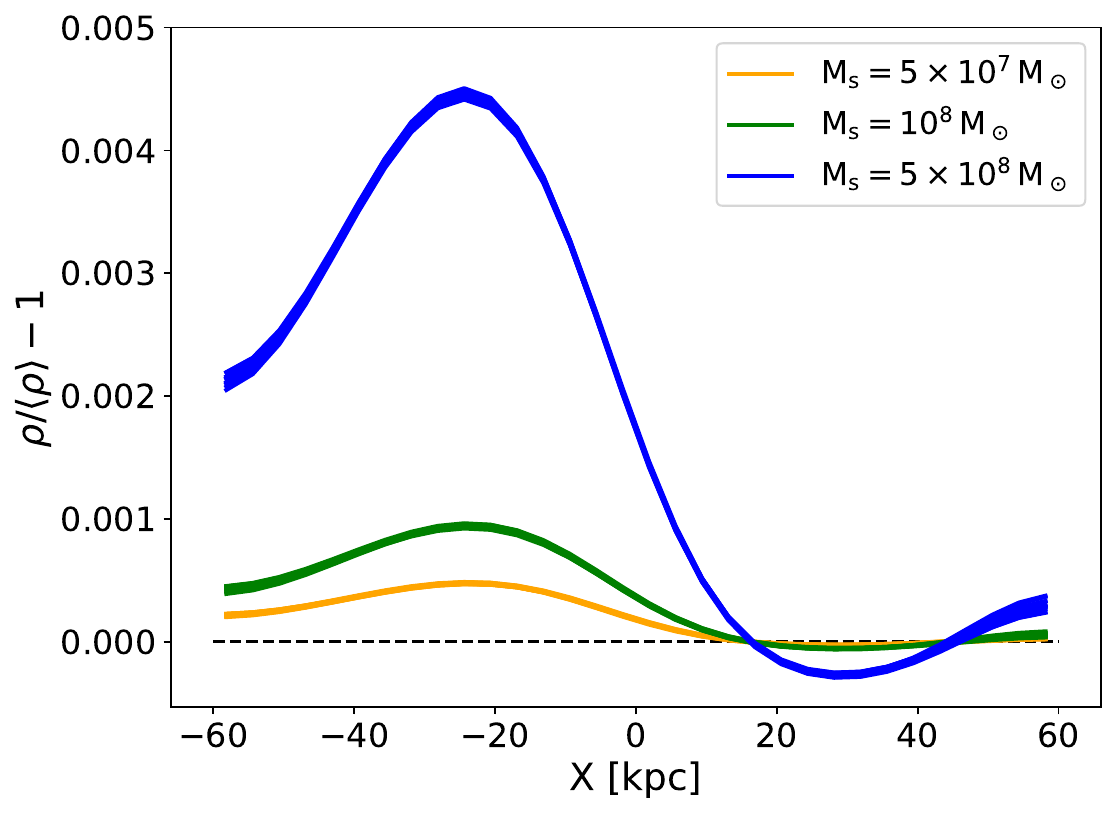}
        \caption{Background subtracted overdensity response profiles averaged across $Y \in [-20, 20] \rm \, kpc$. Each coloured band corresponds to a different subhalo mass and consists of profiles from 12 simulations each with a different initial random seed. The figure demonstrates that the amplitude of the density response scales with the subhalo mass.}
    \label{fig:density_bkg_response}
\end{figure}

We can see how the mass of the subhalo affects the maximal overdensity response in the stellar medium by running the simulation with identical initial conditions both with the subhalo and without. In Figure \ref{fig:density_bkg_response}, we show the Gaussian-smoothed density response of different mass halos after having subtracted the background-only simulation fluctuations exactly from the halo case. We observe that the density peak scales with the subhalo mass and as such we expect the signal to be considerably lower as we explore the detectability of masses lower that $5 \times 10^8 \rm \, M_\odot$. Interestingly, while we observe that the amplitude of the maximum overdensity is a function of the halo mass, we do not see a similar correlation for the relative position of the maximum. In fact, we see the same wake maximum location in the X-coordinate for both the lowest and highest mass halos with the difference being only in the response amplitude. We checked and confirmed that this density peak location is dependent on the subhalo velocity. In particular, we looked at a case where we simulate conditions which mimic the stellar halo at 50 kpc from the Galactic center (Case 2 in Table \ref{tab:params}). In this case, where the perturber is moving $ 25\,\rm  km\,s^{-1}$ slower than in our baseline simulations, we observe that the peak of the density profile is shifted closer to the location of the subhalo.  
Our simulations also hint that the physical size of the stellar wake is considerably larger than what is expected from \cite{Buschmann}. Similar wake characteristics have also been shown in \cite{Foote_2023}. Be as it may, we leave the study of the discrepancy between expected wake size from theory and simulation to future investigations.   

\section{Deep Learning Approach}
\label{sec:ml}

In this section, we introduce our mock data generation procedure and the deep learning model used to detect the stellar wakes caused by subhalos of varying masses. In this first approach, we studied the extent to which we are able to detect a subhalo of any particular mass, formulating the detection as a binary classification problem. 
A given set of $N$ star particles, each described by position and velocity vectors $(\mathbf{p}, \mathbf{v})$ can be described by a $N \times 6$ array $X \in \mathbb{R}^{N \times 6}$.
In general, the ideal discriminator between the subhalo and no subhalo hypothesis is the ratio
$$
D(X) = \frac{L(X | \mathrm{subhalo})}{L(X | \mathrm{no\ subhalo})}
$$
where the likelihoods $L(X | \mathrm{subhalo})$ and $L(X | \mathrm{no\ subhalo})$ are unknown in practice.
We therefore approximate $D(X)$ with the output of a binary discriminator model $\tilde{D}(X)$ that is optimized on simulation samples which implicitly follow the unknown likelihoods.

\subsection{Dataset generation}
\label{sec:datasets}
We used the windtunnel simulations described in Sect. \ref{sec:sim} to generate mock datasets for the purpose of training and evaluating our ML model. In addition to running simulations with a subhalo mass of $5 \times 10^8 \rm \, M_\odot$ (as shown in  \ref{sec:stellarwakes}), we also produced simulations with two additional subhalo mass configurations ($5 \times 10^7 \rm \, M_\odot$ and $10^8 \rm \, M_\odot$), as well as a configuration where no subhalo is present.
We ran the simulation for each mass configuration listed above 48 times using unique random seed settings to draw varying particle velocities from their respective distributions and thus generate additional statistically independent data. 

The full dataset of all simulations is divided into samples, each sample consisting of approximately $1.3 \times 10^6$ star particles, corresponding to 1\% of the number of simulated star particles, $512^3$.
The samples serve as the basis of our analysis, as we aim to distinguish samples from simulations where a subhalo was present with respect to simulations where there was no subhalo.
In a real survey, a single sample could represent a candidate collection of stars of the survey (a region of interest) for which we wish to infer the likelihood of a subhalo being present.

Each sample array $X$ now consists of $1.3\times 10^6 \times 6 \simeq 8 \times 10^6$ values -- the phase-space properties of all star particles.
One approach would be to feed the raw kinematic data of each sample directly to a model for classification.
However, this would result in very large datasets required for model training, and may be nonoptimal due to having to learn from raw data and not inserting any physics priors to speed up the process.
The alternative approach is to define effective observables, computed from the raw star particle kinematics based on a physics \textit{ansatz}.
To summarize the kinematics of a large set of stars, we first start with observables based on 2D histograms by projecting each sample to Cartesian axes in position and velocity.
These projections are produced by equally slicing the simulation box into three slices along the Z-coordinate and binning them into 2D histograms with 32 bins along the x- and y-coordinates.
Based on the star particles in each bin along X and Y, we compute the following four features:
\begin{itemize}
\item bin overdensity with respect to the background density
\item mean speed in the X-Y plane
\item speed dispersion in the X-Y plane
\item velocity divergence in the X-Y plane 
\end{itemize}
The feature histograms for a particular simulation are shown in Fig. \ref{fig:simout}.
After slicing and binning, each sample is thus defined by $3 \times 4$ channels such that each sample is reduced from $8 \times 10^6$ raw observables to $32 \times 32 \times 12 \simeq 12 \times 10^3$ effective observables.

Before training, we used the Gaussian filter from SciPy's \texttt{ndimage} module \citep{2020SciPy-NMeth} on our samples. This filter is designed to smooth the value of each pixel by an amount which is based on the values of its neighbouring pixels. While this smoothing effect blurs the image removing sharp edges, it also works to reduce the overall noise. In our case, we found that a Gaussian kernel of 3 helps to reduce the Poisson noise in the histograms while also preserving the most important features of the wake. The effect that this filter has on the underlying histograms is visible in Fig. \ref{fig:simout}. 

During each training session, we adopted a split of 50\%/33\%/17\% to divide the simulation data into statistically independent training, validation and testing sets. The training set was used for optimizing the model, the validation for the hyperparameter tuning, and the testing set for the final results. For a particular target mass case, we then have 2400 training samples, 1600 validation and 800 testing samples. The derived ML dataset used in our work can be found \href{https://zenodo.org/records/12721089}{in zenodo}.

\subsection{Binary classifier}
\label{subsection:binary_CLSF}

In order to learn the difference between background and subhalo perturbed images, the model has to be provided adequately labelled data to train on. We adopted the simplest labelling possible, where samples derived from simulations containing a subhalo were given an integer label of '1', whereas background simulations (no subhalo) were assigned a label of '0'.
As our physics-based observables are in the form of 2D histograms or equivalently images, Convolutional Neural Networks (CNNs or convnets) are a natural first choice for the model.
The CNN has found wide use in most computer vision domains and has been a major contributor to the rise in popularity of deep learning methods in the past decade \citep{chollet2021deep}.
In our case, we are dealing with images of $32 \times 32$ bins (pixels), with 12 features (channels) per pixel, as described above.
As the dataset generation is based on a complex N-body simulation and we are limited by computational budget, our training dataset consists of only a few thousand samples, putting us in a small dataset regime.
For this reason, we adopt methods that are specifically developed for image classification based on small datasets.
In particular, we use Harmonic Networks~\citep{harmonic_networks}, which use a windowed discrete cosine transform (DCT) to perform a harmonic decomposition of the input features and thus reducing the sensitivity to input noise. 

The harmonic layer is different from a standard convolutional layer as it does not learn filters for extracting spatial correlation, but instead operates in the frequency domain and learns the weights of the DCT filters.
According to the work presented in \cite{harmonic_networks_limited_data}, these layers perform better in the case of small datasets when compared to traditional CNNs, which we have confirmed in our dataset directly.

The model was trained with the Adam \citep{kingma2017adam} optimizer and binary focal cross entropy loss function~\citep{lin2017focal} to give larger weight to hard-to-classify samples. 
The model was implemented using Keras \citep{chollet2015keras} and TensorFlow \citep{tensorflow2015-whitepaper}.

\begin{figure}[t]
    \centering
        \includegraphics[width=0.45\textwidth]{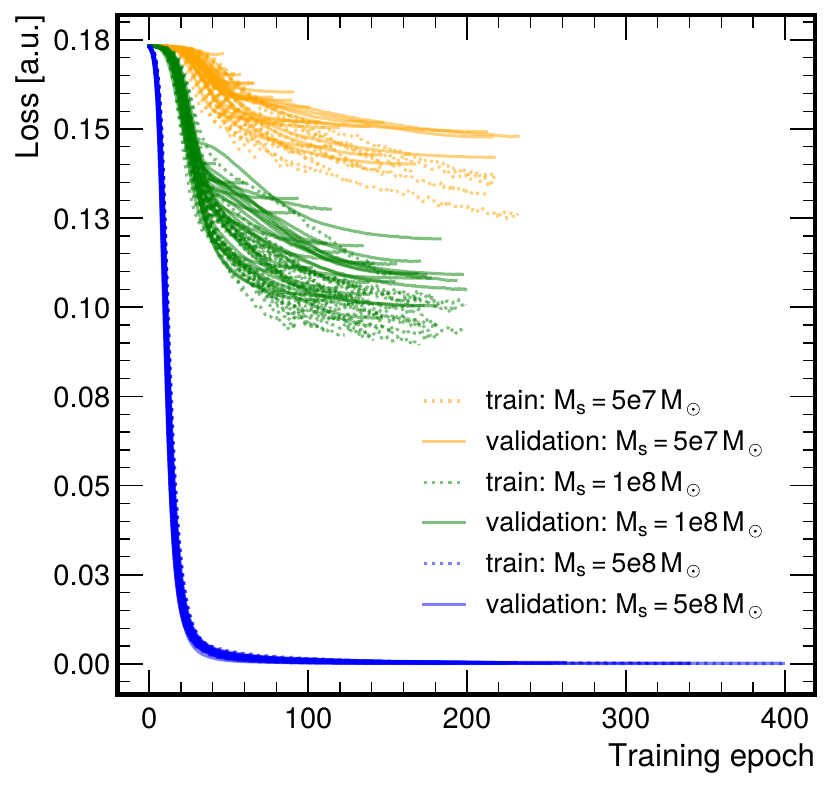}

    \caption{Training and validation loss of the binary classifier model after running the model 30 times. The training loss of the model measures the discrepancy between the predicted outputs of the model and the actual targets in the training dataset. The validation loss depicts the model performance on the validation data and is therefore a measure how well the model generalizes to unseen data.} 
    \label{fig:CLSF_training}
\end{figure}

The choice of the exact architecture and the number of layers and filters per layer was based on hyperparameter tuning.
We used the RandomSearch in the KerasTuner \citep{omalley2019kerastuner} framework for parameter tuning.
The scanned hyperparameters, their initial ranges, and the final values are shown in Table \ref{table:hyperparameters}, along with the evolution of the loss in Fig.~\ref{fig:CLSF_training}. As we divided our simulation box into three slices in the Z-coordinate, the 'number of z-slices' in Table \ref{table:hyperparameters} refers to how many of these slices we included in the training. Similar to traditional 2D convolutional layer, the 'filters' hyperparameter configures the output dimension of the layer. In the table, we show the output dimension of the first layer which we increase two-fold after each successive Harmonic layer. 'Learning rate', 'dropout' and 'activation' correspond to the step size of the loss function, the amount of regularization used after each layer and which activation function we used. We also show the explored range of kernel sizes for the first two Harmonic layers. In addition to this, we experimented with adding additional layers to these baseline layers showing this as 'extra layers' in the table. Finally, the last parameter in the table is a scalar factor with which we expand the output dimensionality of the second to last fully connected layer in our model. 
The full hyperparameter tuning took about 20 hours on one Nvidia RTX2070S.

\begin{table}[ht]

\centering
\begin{tabular}{lcl}
\hline
\textbf{Hyperparameter} & \textbf{Range} & \textbf{Final Value} \\ \hline
number of z-slices & [1, 2, 3] & 3 \\
filters & [4, 128] & 32 \\
learning rate & [1e-8, 1e-2] & 1.9602e-06 \\
dropout & [0, 0.6] & 0.49259 \\
activation & [relu, selu] & relu \\
kernel of 1st layer & [3, 10] & 9 \\
kernel of 2nd layer & [1, 3] & 2 \\
extra layers & [0, 3] & 1 \\
filter expansion & [1, 16] & 2 \\ \hline
\end{tabular}
\caption{Hyperparameter selection for our binary classifier.}
\label{table
}
\label{table:hyperparameters}
\end{table}

With the aim of producing statistically independent simulations and training samples, we adopted a unique random seed every time we drew the initial conditions before running any simulation. Due to fluctuations in the simulations, we expect variability in the training performance. In order to assess the effect that this has on our model's performance, we train the model 30 times for each adopted subhalo mass target. Every run we pick a random permutation of train, validation and testing samples such that the sets of their origin simulations using seeds $k$, $l$ and $m$ obey $k_{train} \cap l_{val} \cap m_{test} = \emptyset$. This allows us to separate training and testing samples during training, average any metrics relevant to the model performance across the training runs, and also report the error bars. 

During each training run, we use early stopping to halt training after validation loss has not decreased during the last 5 epochs. With a constant learning rate of $\simeq 2 \times 10^{-5}$, the total training time for a particular mass case (30 runs) adds up to about 1 hr on one Nvidia RTX2070S.
We show the training and validation loss progression in Fig. \ref{fig:CLSF_training}, where each set of coloured lines corresponds to a particular subhalo target case. As expected, the approximate final loss value plateaus differ for each subhalo mass target and we see that the model's training difficulty decreases as the subhalo mass increases.
We also observe that the final training and validation loss values exhibit more scatter in the case of the lighter subhalo masses. When using datasets with a subhalo of $5 \times 10^8 \rm \, M_\odot$, the training is much more stable as both losses plateau at smaller values and show much smaller variance between training runs. 

We present our binary classifier's final results and discuss its detection performance in Sect. \ref{sec:results}.

\section{Results}
\label{sec:results}

\subsection{Effect of spatial and kinematic training features}

\label{subsection:effect_of_features}

\begin{figure}[t]
    \centering
        \includegraphics[width=0.45\textwidth]{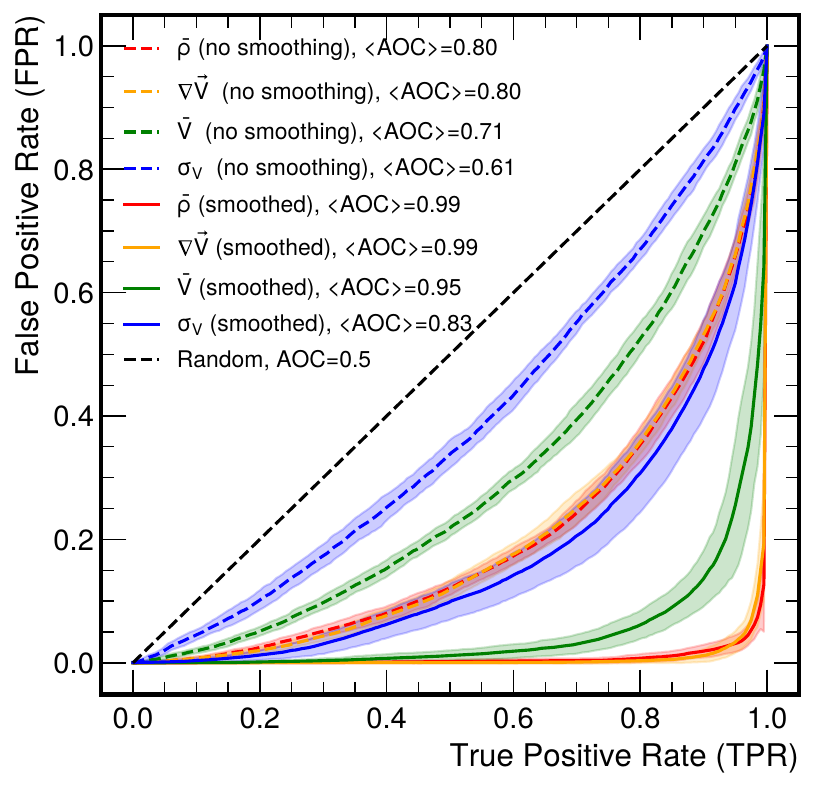}
        \caption{Binary classifier performance for $5 \times 10^8 \rm \, M_\odot$ when training on images generated from the middle slice ($Z \in [-20, 20]$ kpc) of the simulation box. Different colored bands depict the performance when training on different features such that: red - overdensity, yellow - divergence, green - mean speed, blue - speed dispersion. Model performance where training was done on Gaussian smoothed features is depicted by the solid lines, whereas dashed lines show when training was done on unsmoothed features. We observe that best performance is achieved by using smoothed features out of which overdensity and velocity divergence are most effective.}
    \label{fig:ftrs_effects}
\end{figure}

We used the binary classifier model described in Sect. \ref{subsection:binary_CLSF} to study which physical observables or their combination would be most useful for detecting the stellar wakes. We quantified the performance using the Receiver Operating Characteristic (ROC) curves and the Area Over the Curve (AOC). The ROC curves represent the model's sensitivity (true positive rate, TPR) and specificity (false positive rate, FPR) across all possible threshold settings. To assess the physics and model performance, the model was trained and evaluated 30 times on independent sets of training and testing datasets. Below we summarise results from testing different feature engineering and selection options.

\begin{itemize}
\item \textbf{Gaussian smoothing:} We inspected how the model performance is affected when the Gaussian filter (introduced in Sect. \ref{sec:datasets}) is applied to the training features. Fig. \ref{fig:ftrs_effects} depicts performance for the target mass $5 \times 10^8 \rm \, M_\odot$ when training our baseline binary classifier model (detailed in Sect. \ref{subsection:binary_CLSF}) separately on each of the four features introduced in Sect. \ref{sec:datasets}. Solid lines show results when training is done on Gaussian smoothed features whereas dashed lines show results when smoothing is turned off. We found that in the case of smoothed features, we see a significant improvement ($\approx 25-35$\%) in model performance when compared to their non-smoothed counterparts.
\item \textbf{Individual feature performance:} From the same figure, we observed that overdensity and divergence (AOC = 0.99 for both) seem to be the most effective training features  followed by mean speed (AOC = 0.95) and lastly speed dispersion (AOC = 0.83). We repeated the same exercise for a lower mass target case ($10^8 \rm \, M_\odot$) to see whether these conclusions are affected by the mass of the simulated subhalo but our results remained qualitatively the same. Namely, in terms of AOC values, mean dispersion yields 0.58, mean speed in X-Y 0.68 (17.24\% increase) and divergence yields a value score of 0.73 (a further 7.35\% increase). We also checked that by using the Cartesian velocity component $v_x$ instead of the mean speed ($v_{xy}$), there is no statistically significant difference in performance between the two.

\item \textbf{Combining kinematic features:} We studied how our model performance is affected when combining different kinematic features. For this purpose, we performed three training runs: first we trained only on divergence, then added dispersion, and finally including mean speed. This enabled us to quantify the difference between performance when using one, two or three kinematic features. We observed AOC values of 0.73, 0.71 and 0.71 respectively. While all features exhibit positive constraining power when used individually, we did not observe a stacking effect in the overall performance of the model when other kinematic features are combined with divergence. We concluded from these results that divergence, as expected, already contains much of the information present in the other two features.

\item \textbf{Combining kinematic and spatial features:}
We found that training on overdensity and velocity divergence yields the best classification performance, with no significant improvement when adding the other 2 kinematic features.
For this 2-feature combination and the $10^8\,\rm M_\odot$ mass case, the model correctly identifies 74\% of signal samples at optimal threshold while misclassifying background samples at a rate of 35\%. In contrast, using only overdensity information results in 70\% and 40\%, respectively This  highlights the added value of kinematic information in distinguishing signal from background.
\item \textbf{Miscellaneous feature experiments:} In addition to the above, we also tried a few other options with hopes to increase model performance. For example, we looked at using relative velocities (normalizing the kinematic features around 0 analogously to Eq. \ref{eq:overdensity}). Furthermore, since we slice our data into three equal slices in the Z-coordinate, we also investigated whether we should keep or exclude the outer layers during training. Using small experiments we confirmed that adopting relative velocities instead of absolute ones and the inclusion of the upper and lower Z-slice do not improve our results. Consequently, we decided to keep absolute values and proceeded with training only on data from the middle slice of the box, which contains the majority of the wake.
\end{itemize}

Our findings described above may hint that the detection of subhalo masses considered in the current work is achievable with either positional or kinematic data but a combination of both yields strongest results.

\subsection{Binary classification performance}
\label{subsec:binary_CLSF_perf}

\begin{figure}[t]
    \centering
        \includegraphics[width=0.45\textwidth]{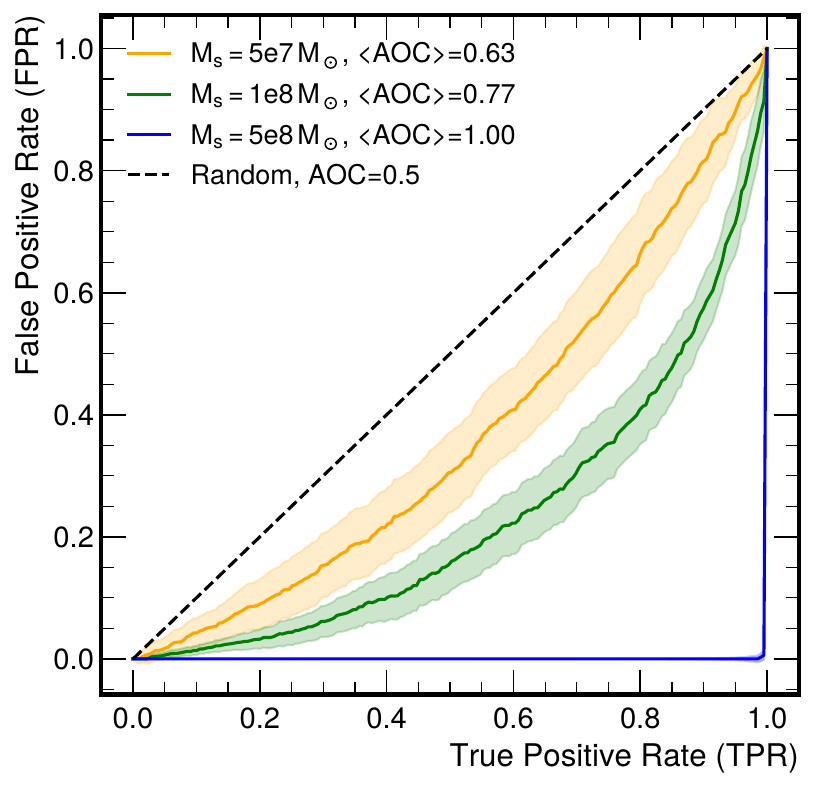}
    \caption{Receiver Operating Characteristic (ROC) curves for the binary classifier trained on datasets with varying subhalo masses. The curves represent subhalos with masses $5 \times 10^7 \rm \,M_\odot$ (<AOC>=0.63), 
$10^8 \rm \,M_\odot$ (<AOC>=0.77), and $5 \times 10^8 \rm \,M_\odot$ (<AOC>=1.00) solar masses. The width of the bands represents the standard deviation of the curves when training and evaluating the model 30 times. The median area over the curve (<AOC>) values indicate the classifier's performance in distinguishing between background (no subhalo) and the presence of a subhalo. The performance of the binary classifier scales with the mass of the subhalo.}
    \label{fig:CLSF_ROC}
\end{figure}

We present the performance of our binary classifier for our chosen target cases in Fig. \ref{fig:CLSF_ROC}. For a particular target case, we show the median ROC (solid lines in Fig. \ref{fig:CLSF_ROC}) as well as the standard deviation across multiple training runs (shown as the shaded area in Fig. \ref{fig:CLSF_ROC}).
We see that at all tested masses, the model is able to distinguish between samples from the background and subhalo simulations better than random choice.
Furthermore, as expected, subhalos with higher masses and thus a more prominent wake are detected with higher accuracy.
This demonstrates that there is sufficient residual information to distinguish the presence of a subhalo in the wake of the stellar particles down to $M = 5 \times 10^7 M_\odot$ under the ideal conditions.
As was hinted already in the training loss curves of Fig. \ref{fig:CLSF_training}, we see that for  $M = 5 \times 10^8 M_\odot$, the scatter appears to be negligible across the runs while the same cannot be said about the other target cases.

The variance in the ROC scatter and median AOC was investigated in dedicated ablation studies by changing the size of the training dataset. For the lower-mass subhalo target of $5 \times 10^7 \,\rm M_\odot$, the binary classifier's performance, trained on 25\%, 50\%, 75\%, and 100\% of the available data, resulted in AOC values of $0.586 \pm 0.118$, $0.609 \pm 0.083$, $0.621 \pm 0.064$, and $0.628 \pm 0.059$, respectively. In addition, we also investigated how architectural changes from our baseline model impact our results and found that the current configuration is optimal within the tested set of configurations. These studies confirmed our hypothesis that our results are most significantly affected by the amount of available training data. That is, by increasing the number of statistically independent samples, the training becomes both more stable and accurate. 

The remedy for this issue might seem trivial (i.e. generate more data), but in practice, running more simulations after a certain point becomes cumbersome as it would require implementing data reduction techniques or access to substantial computing resources. In recent years, the development of deep generative models and emulators have begun to push boundaries in terms of fast data generation for different simulation based inference problems (see e.g. in \cite{ramesh2022gatsbi, Hemmati_2022}) which may be interesting to explore in future studies. We expect that with increased simulation datasets our results can be significantly improved, while at the same time, the effect of uncertainties on the detectability can be studied.

\subsection{Multiple mass hypothesis testing}

\begin{figure}[t]
    \centering
        \includegraphics[width=0.45\textwidth]{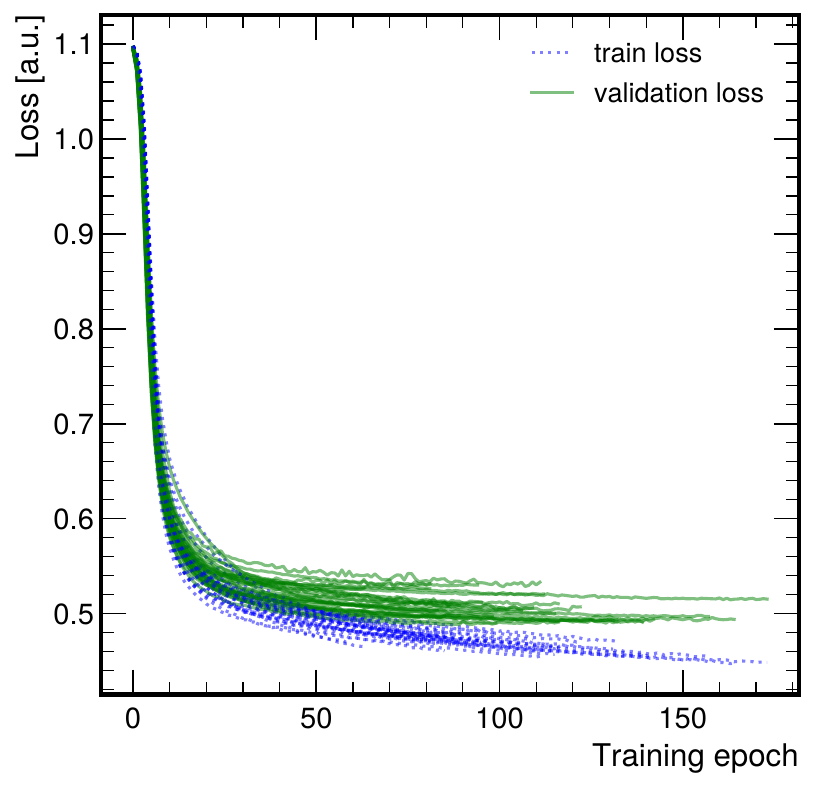}

    \caption{Training (blue) and validation (green) loss of the Multiple mass hypothesis classifier model after running the model 30 times with early stopping.} 
    \label{fig:multiclass_training}
\end{figure}

In addition to studying the ability to infer the presence of a subhalo in our samples, we also investigated how well we can discern between the different subhalo mass cases in a multiple-hypothesis case. This time, instead of using both background and signal samples, we trained exclusively on samples containing all three signal cases labelling them from lowest mass to highest as 0, 1 and 2. Like before, we trained the model with early stopping 30 times where at the start of each run we picked a random permutation of simulation seeds for training, validation and testing. Figure \ref{fig:multiclass_training} shows the training and validation loss curves for all training runs. We see that for each iteration of the dataset shuffle, both training and validation losses decrease smoothly over time and start to plateau at around 100 training epochs.

\begin{figure}[h]
    \centering
        \includegraphics[width=0.45\textwidth]{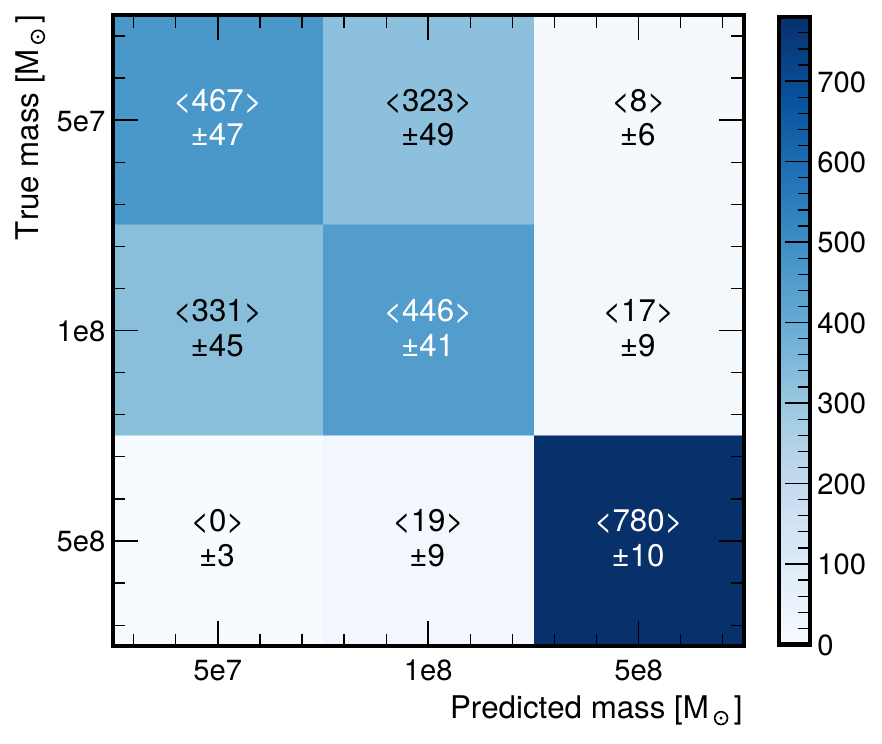}

    \caption{Multiple mass hypothesis performance summarised in terms of the predicted and true mass of the subhalo in test samples. Each element in the confusion matrix is characterised by the mean number predictions and the standard deviation across 30 runs of training and evaluating.} 
    \label{fig:multiclass_performance}
\end{figure}

Instead of a single prediction score, each test sample is given three scores each of which represent the probability of belonging to a particular target class. In each of the cases, the model is able to discern between samples in the testing dataset when there is a clear difference between the prediction distribution of the samples actually belonging to the particular target case with respect to the rest. We can then summarise the accuracy of our model i.e. how well it is able to discriminate between these distributions with a confusion matrix in Fig. \ref{fig:multiclass_performance}. Ideally we would like to maximize the values of elements on the main diagonal which depict the number of instances where the model is able correctly predict the mass of the subhalo. The off-diagonal elements show mismatches between predicted and true labels and thus indicate which targets are harder to classify for the model. Since we run the model 30 times, we show the prediction count values of each element in the matrix by computing the mean and standard deviation across all runs. We note that since we average many training runs, we do not expect the counts across columns to sum to the total number of samples (800) in each target mass test dataset. We do however expect this sum to be within the standard error across the runs.   

Much like we saw in the binary classifier analysis of Sect.~\ref{subsec:binary_CLSF_perf}, we again see that the model performs best in the case of the heaviest subhalo mass ($5 \times 10^8 M_\odot$). In this case the model was able to identify the correct mass of approximately 780 samples with a small scatter in the mean number of predictions ($\pm 10$) and mislabel 20 samples as other targets. For the lower masses of $M = 5 \times 10^7 M_\odot$ and $M = 10^8 M_\odot$, the task was more challenging as we observe a larger scatter in correct predictions counts ($\pm 47$ and $\pm41$) as well as a tendency to mislabel the samples between these two. In both cases, about 300 samples were mislabelled. Since wake effects created by a subhalo of mass $M = 5 \times 10^7 M_\odot$ are considerably smaller than those created by $5 \times 10^8 M_\odot$, similar performance between the lower mass cases points to a difficulty in identifying intermediate mass samples.

\subsection{Detection performance 50 kpc from the Galactic center}

On top of inspecting the subhalo detection performance at 30 kpc from the Galactic center we also looked at what would happen if the perturber was in orbit at 50 kpc which is roughly the distance to the LMC. While \cite{Foote_2023} studied wakes created by LMC-sized subhalos, in this study we are interested in effects created by much smaller subhalos. We then ran our simulation again using our intermediate subhalo mass, $M = 10^8 \rm \, M_\odot$, and like before, configured the background and perturber phase-space parameters to literature informed values (summarised in Table \ref{tab:params}). By modulating these different parameters we expect the density response of the perturber to change. For example, the Chandrasekhar dynamical friction equation \cite{Chandrasekhar} hints that at a constant perturber velocity, the reduction in the ambient velocity dispersion will result in a larger deceleration (i.e. density wake) of the perturber. This classical dynamical friction equation however does not take into account the effects of self-gravity, and applies only to specific idealised conditions. Furthermore, the combined effects of all background and subhalo parameter changes (e.g. subhalo velocity, stellar and DM mass density, etc.) on the actual amplitude and extent of the stellar wake are not easily estimated beforehand and are thus interesting to explore. We leave a full investigation of the relationship between simulation phase-space parameters and wake observables for a future work and continue with results from our ML analysis.  

\begin{figure}[h]
    \centering
        \includegraphics[width=0.45\textwidth]{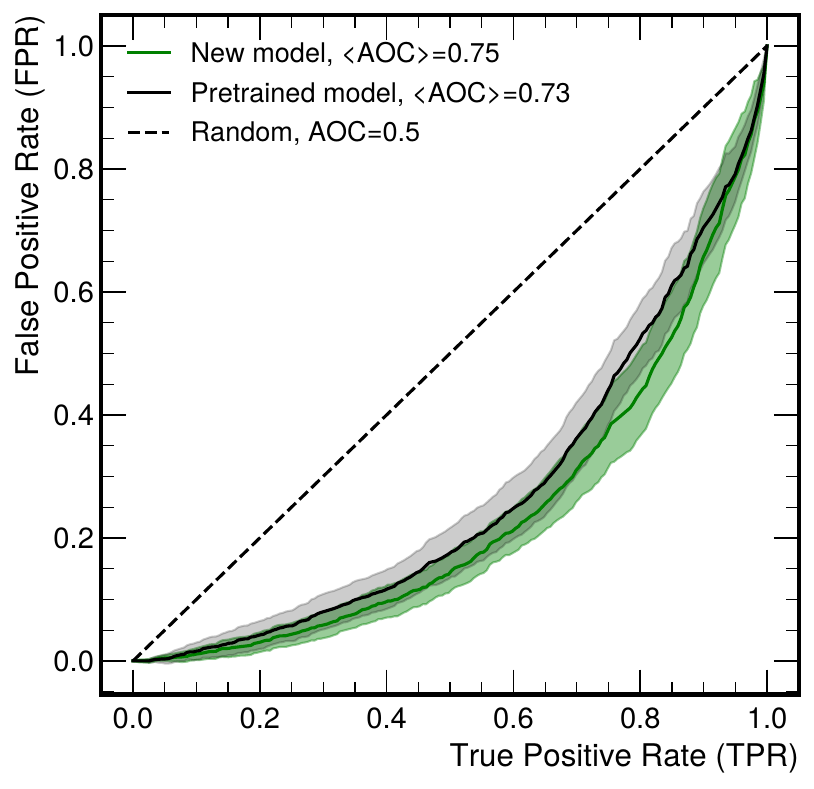}

    \caption{Binary classification results of $M_s = 10^8 \rm \, M_\odot$ when orbit is placed at 50 kpc from the Galactic center. Green shows performance of new model which is trained and evaluated on simulation data from 50 kpc. The black line shows the  model performance when training is done on data describing a subhalo orbit at 30 kpc from the Galactic center.} 
    \label{fig:Binary_CLSF_50kpc}
\end{figure}

Using data from the 50 kpc simulations, we derived new ML samples in exactly the same way as was described in Sect. \ref{sec:datasets}. This way we ensure that the performance comparison between these two cases is done on a fair basis. Without making any changes to the binary classification model, we trained the model again with the same setup and we summarise the results from these runs in Fig. \ref{fig:Binary_CLSF_50kpc} as the green band. We see similar performance to the first case for these new samples. This shows that our binary classification model is able to learn from a completely new and independent dataset and that our previous results are not case specific. One physical interpretation of the similarity between the two cases could be that 20 kpc is too small a distance for the phase-space parameters to change enough to have an impact on our detection model. In other words, the slopes of e.g. mass density and velocity dispersion profiles are too small and perhaps the Galactocentric distance should be even larger. The usefulness of (small mass) subhalo simulations in environments out to > 100 kpc is another question as the lack of stellar observations with adequate precision discourages the detection of the subhalo induced wake effects.

In addition to the above, we also looked at the detection performance when evaluating the new data (subhalo orbit at 50 kpc) on a previous model that was trained on data when the subhalo was 30 kpc from the Galactic center. We show the results as the gray band on Fig. \ref{fig:Binary_CLSF_50kpc}. We again see similar performance as before which is a good indication that our model is able to generalise to new conditions. Expectedly, the AOC of the black ROC curve is smaller as in the case of the green band the model was trained on the new dataset and is therefore better tailored to make predictions on it.   

\section{Discussion}
\label{sec:discussion}

The physical setup of the idealised simulations described in this work can certainly be improved upon in many aspects. For example, the current setup does not include the gravitational potential of the Galaxy or the effect of tidal stripping. Also, it would be interesting to see how the stellar wakes and their detection performance is affected when using different density profiles (e.g. Navarro-Frenk-White, Einasto, etc.) for the subhalo. 

In a future study, we also plan to investigate how our results are affected by the inclusion of observational effects. Specifically, we would like to relate our data from ideal simulations to real surveys (e.g. Gaia \cite{Gaia_mission}) by studying the detectability at varying error levels in an observational frame of reference. Due to their large spatial extension, we do not expect to detect stellar wakes in their entirety. However, we know from \cite{BAZAROV2022100667} that we are able to observe a signal when looking at regions near subhalos on a star-by-star basis. In any case, acquiring intuition on the actual physical scales of the stellar wake phenomena could be an important step as we start look for and identify suitable regions of interest from real survey data.

In addition to creating mock datasets in a new frame of reference, the ML models will also need to be adapted. In the current work, we used three overdensity images (slices) per sample for training. This means that in order to evaluate already trained models with new data, the input needs to conform to the same dimensionality that is (N, 32, 32, 2). Creating similar Z-slices in an observational setting is not as straightforward as was the case in our idealised box simulation setup. In our case, the thickness of the slices is chosen arbitrarily to divide the simulation region into three equal slices wherein the middle layer contains the subhalo and majority of the stellar wake. In a mock dataset of an observational region of interest, the decomposition of data into slices along the line of sight might not be justified altogether as the position of subhalo is not localized and the direction of motion is arbitrary with respect to the coordinate axes.

Even though we have achieved very good classification performance on samples containing very heavy subhalos (i.e. $M_s > 10^8 \rm \, M_\odot$), we still have room for improvement in identifying lower mass target cases. One direction to tackle this would be to consider alternative ML approaches and architectures as the models described in the current work are certainly not exhaustive. For example, it would also be interesting to see how well one would be able to predict the subhalo mass as part of a regression model setup.

While other methods are possible, we found that in our case, the key limiting factor is the amount of available training data. By retraining our binary classifier while modulating the amount of available training data we have seen that with more data we achieve increased AOC values and smaller scatter in the ROC curves. The data problem is something that could be overcome with access to considerably larger computing resources or finding alternative ways to generate simulation data faster (e.g. emulators, generative models, etc.). Since we have not reached a performance plateau, it is difficult to give estimates on sufficient training dataset sizes. We would like to note that the ultimate goal, which is the focus of a future study, is how well our model generalises to observational data rather than trying to learn the simulation data perfectly.

\section{Conclusions}
\label{sec:conclusions}

Constraining the SHMF in the sub-galactic mass regime is an important endeavour as we aim to understand more about the particle nature of DM. Theoretically predicted dark subhalos are extremely difficult to detect as their presence can only be inferred from gravitational effects on the surrounding stellar medium. In this paper, we studied the strength of the DM subhalo induced gravitational signal by investigating how well we can detect individual stellar wakes induced by orbiting subhalos in the stellar halo.

We implemented windtunnel simulations with self-gravity enabled using PKDGRAV3, replicating the ambient phase-space conditions of DM and stars at 30 and 50 kpc from the Galactic center. Interestingly, we observed stellar wakes in line with those for larger perturber masses as described in \cite{Foote_2023}, but significantly more spatially extended than those in \cite{Buschmann}. The former study finds that, for perturbers with masses of $\mathcal{O}(10^{11}\,\rm M_\odot)$,  the inclusion of self-gravity increases the magnitude of the density response by roughly 10\% while also significantly extending the length of the overdensity and kinematic wake. In the latter work, self-gravity was not considered, but we found in our simulations that although the removal of self-gravity reduces the spatial extension of the wake, this omission alone does not fully account for the difference.

Then, we derived mock datasets by binning the simulated data into 2D histograms and computing different physical observables in each bin to be used as training features. The phase-space features that we implemented were the overdensity, mean speed on the X-Y plane ($V_{xy}$) and its dispersion and divergence. We found that by applying a Gaussian smoothing filter on the features prior to training, we see a significant increase in classification performance. Even though all considered features showed non-trivial constraining power when used exclusively, we found that the combination of overdensity and velocity divergence is equivalent to using all four. This became evident as including additional kinematic features did not significantly improve classification performance when divergence was already included in the training dataset. In any case, these findings hint that stellar wakes may be best found in ongoing or future stellar surveys by using a combination of positional and kinematic information which in our study exhibited comparable constraining power.

Finally, we divided our ML approach into two. First, we investigated how well we are able to infer the presence of different mass subhalos in the generated images. We implemented a binary classification model which we then trained and evaluated on our three target mass cases:  $5 \times 10^7 \rm \, M_\odot$, $10^8 \rm \, M_\odot$ and $5 \times 10^8 \rm \, M_\odot$. We saw that for all the chosen target cases we are able to infer the presence of a subhalo at a rate which is better than random. Expectedly, we observed that the performance follows a hierarchical trend such that more massive subhalos exhibit more signal and are easier to detect. We also investigated our binary classification model's performance having simulated a subhalo of mass $10^8 \rm \, M_\odot$ at 50 kpc from the Galactic center. Using this new simulation data, we compared the classification performance of a model that was trained on a newly derived ML dataset against the pretrained model at 30 kpc. We found similar results in both cases and saw that our model's performance is generalisable to data from simulations with different physical conditions.

We also studied classification between different subhalo masses in a multiple-hypothesis case. We find that the model is able to recognize and correctly label subhalos of mass $5 \times 10^8 \rm \, M_\odot$ about 97\% of the time, demonstrating a potential capability to constrain subhalo masses.\\

The current work is summarized as follows:
\begin{itemize}
    \item We use machine learning to evaluate how effectively we can detect {\it individual} stellar wakes induced by DM subhalos in the MW's stellar halo.
    \item Our simulated stellar wakes are in line with \cite{Foote_2023}, but significantly more spatially extended than previously reported in the literature. We found that the inclusion or omission of self-gravity does not fully account for the difference.
    \item In the context of detection performance, we found that: 
    \begin{itemize}
    \item Gaussian smoothing plays a crucial role, improving AOC values by approximately 25-35\%.
    \item The combination of overdensity and velocity divergence results in maximal performance, achieving a TPR of 60\%/74\%/99\% and an FPR of 41\%/35\%/1\% for $5\times 10^7\,\rm M_\odot$/$10^8\,\rm M_\odot$/$5\times 10^8\,\rm M_\odot$ mass cases.
    \item Training only on overdensity reduces the performance to a TPR and an FPR of 70\%/97\% and 40\%/5\%, respectively, for the $10^8\,\rm M_\odot$/$5\times 10^8\,\rm M_\odot$ subhalo cases.
    \item
    With the amount of training data available (4800 samples), the $5 \times 10^8 \rm \, M_\odot$ subhalo is perfectly identifiable while using only a 1\% fraction of all star particles present in the snapshot (i.e. 1.3M star particles).

    \item Detection performance for smaller subhalos is significantly reduced, with the amount of available training data being the key limiting factor.

    \item We found that our performance remains effectively unchanged when varying the subhalo's position relative to the Galactic center within 50 kpc, demonstrating generalizability to data under different physical conditions.
    
    \item In a multi-class classification scenario, the model performed best for the heaviest subhalo mass ($5 \times 10^8 \rm \, M_\odot$), correctly classifying around 97\% of these samples.
 \end{itemize}   
\end{itemize}

\begin{acknowledgements}

We would like to express our gratitude to the referee for their constructive feedback, which has significantly contributed to the improvement of the results presented in this work. We thank Hayden Foote for valuable discussions. We would like to thank Martti Raidal for pointing us towards using ML approaches to study dark subhalo detection in the MW. This work was supported by the Estonian Research Council grants PSG938, PSG864 and PRG1006, the Estonian Ministry of Education and Research (grant TK202) and the European Union's Horizon Europe research and innovation programme (EXCOSM, grant No. 101159513).
We further acknowledge the support of the European Consortium for Astroparticle Theory in the form of an Exchange Travel Grant. I.A.A. and C.D.V. thank the Ministerio de Ciencia e Innovación (MICINN) for financial support under research grant PID2021-122603NB-C22. The results and figures presented in this work were made possible thanks to the following software libraries: Matplotlib \citep{matplotlib}, NumPy \citep{numpy}, SciPy \citep{2020SciPy-NMeth}, HDF5 for Python \citep{collette2013python} and Jupyter \citep{jupyter}. This research was supported by an academic grant of an A100 GPU from Nvidia. 

\end{acknowledgements}

\bibliographystyle{aa}
\bibliography{main}

\begin{appendix}

\section{$V_x$ and $V_y$ velocity maps}

\begin{figure}[h]
    \centering
    \begin{subfigure}[t]{0.4\textwidth}
        \centering
        \includegraphics[width=\linewidth]{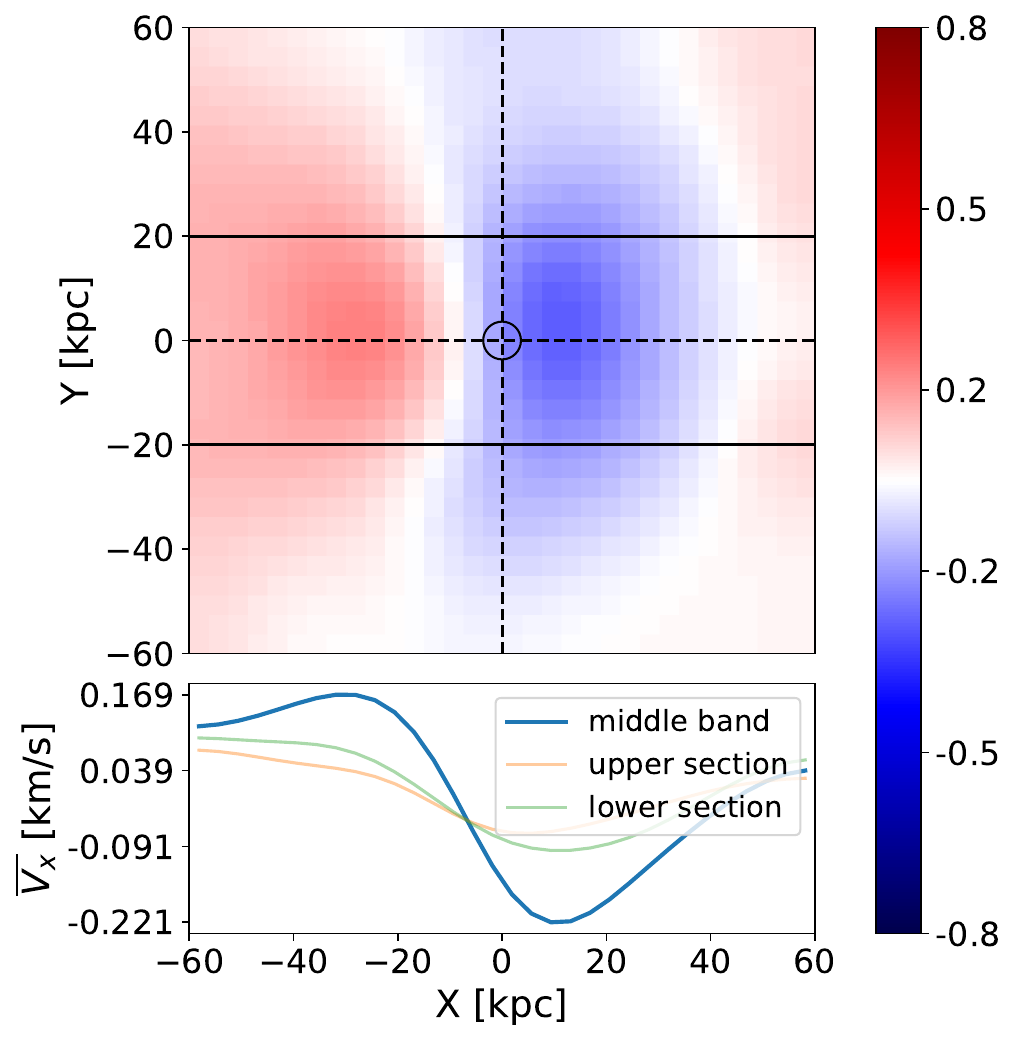} 
        \caption{Mean $V_x$ velocity [km/s]} 
    \end{subfigure}

    \vspace{1cm}
    \begin{subfigure}[t]{0.4\textwidth}
        \centering
        \includegraphics[width=\linewidth]{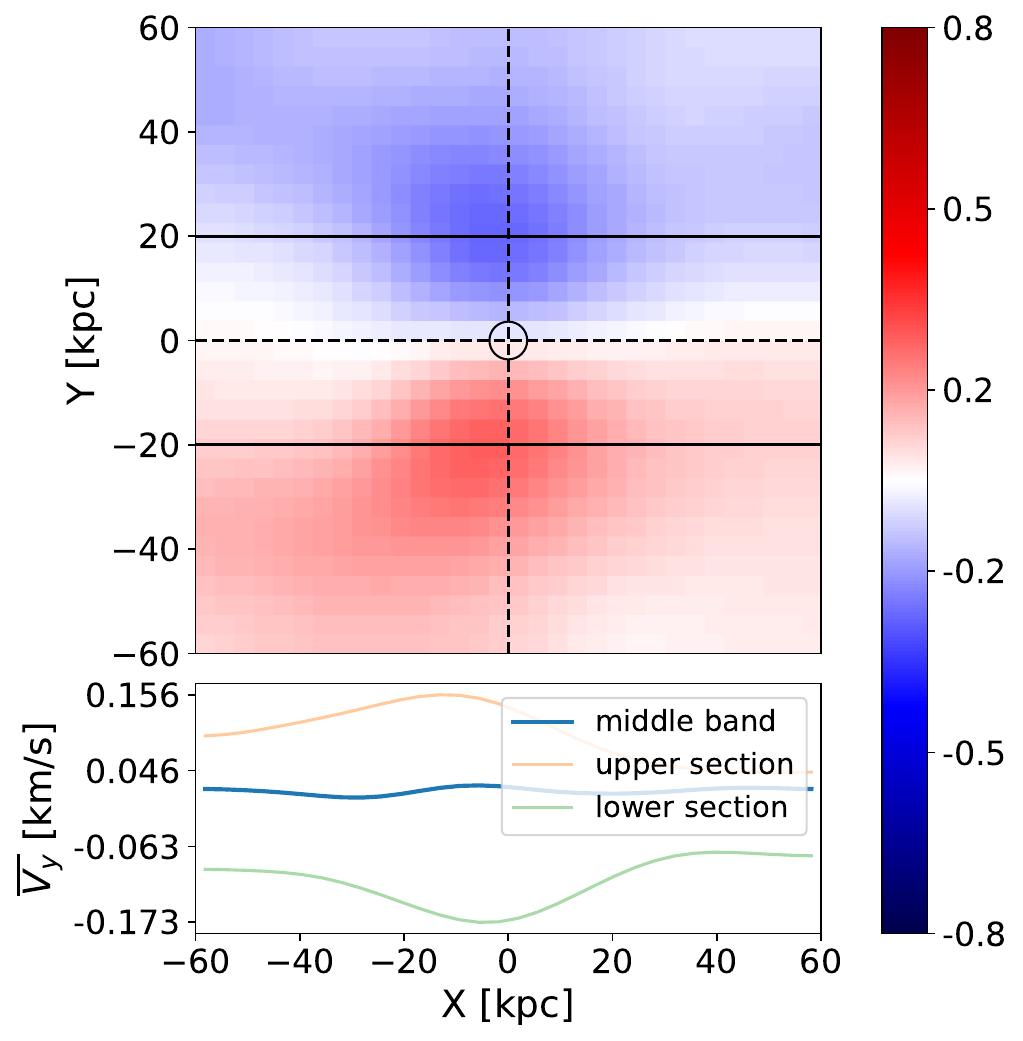} 
        \caption{Mean $V_y$ velocity [km/s]}
    \end{subfigure}

    \caption{Stellar velocity maps of $V_x$ (a) and $V_y$ (b) in a simulation containing a subhalo of mass $5 \times 10^8 \rm \, M_\odot$.  }
    \label{fig:appendix_vx_vy}
\end{figure}

Figure \ref{fig:appendix_vx_vy} shows the $V_x$ and $V_y$ velocity maps of star particles in a simulation containing a subhalo of mass $5 \times 10^8 \rm \, M_\odot$. In the same way as in Fig. \ref{fig:simout} of the main text, stars of z-slice $z \in [-20, 20] \, \rm kpc$ are binned in a 2D histogram with 32 bins on both axes. Inside each bin, the velocity components of the star particles are summed and averaged across ten simulations. In this way, the kinematic signatures of the wake become much more clearer in the figures.

We note that the velocity scales between \ref{fig:timing2} and \ref{fig:appendix_vx_vy} differ due to the fact that in the former we show the mean speed in the X-Y plane whereas in the latter we show maps of the velocity components ($V_x$, $V_y$) separately. Since the stellar velocities ($V_x$, $V_y$, $V_z$) are drawn from distributions centered on 0 km/s in the reference frame of the simulation box, the mean velocities in A.1 also naturally average to around 0 km/s.

\end{appendix}

\end{document}